%% file: main.tex
\documentclass[conference]{IEEEtran}
\IEEEoverridecommandlockouts
\usepackage{cite}
\usepackage{amsmath,amssymb,amsfonts,bm}

\usepackage{tabularx}
\usepackage{booktabs}
\usepackage{array}
\newcommand{\heading}[1]{\multicolumn{1}{c}{#1}}

\usepackage{graphicx}

\ifCLASSOPTIONcompsoc
    \usepackage[caption=false, font=normalsize, labelfont=sf, textfont=sf]{subfig}
\else
\usepackage[caption=false, font=footnotesize]{subfig}
\fi
\usepackage{xcolor}

\usepackage{tikz}
\usetikzlibrary{
    calc,
    chains,
    shapes, shapes.geometric,
    arrows, arrows.meta
}

\usepackage{pgfplots}
\pgfplotsset{compat=1.14}
\usepgfplotslibrary{fillbetween}
\usepackage{pgfplotstable}
\pgfplotstableset{col sep=comma}

\usepackage{siunitx}

\definecolor{set1-darkred}{RGB}{215,48,39}
\definecolor{set1-lightred}{RGB}{252,141,89}
\definecolor{set1-verylightred}{RGB}{254,224,144}
\definecolor{set1-verylightblue}{RGB}{224,243,248}
\definecolor{set1-lightblue}{RGB}{145,191,219}
\definecolor{set1-darkblue}{RGB}{69,117,180}

\makeatletter
\newcommand\thefontsize[1]{{#1 The current font size is: \f@size pt with \the\baselineskip baselineskip\par}}
\makeatother

\begin{document}

\onecolumn

\noindent \textcopyright{} 2020 IEEE. Personal use of this material is permitted. Permission from IEEE must be obtained for all
other uses, in any current or future media, including reprinting/republishing this material for advertising or
promotional purposes, creating new collective works, for resale or redistribution to servers or lists, or reuse
of any copyrighted component of this work in other works.

\twocolumn

\title{%
Machine Learning Clustering Techniques for Selective Mitigation of  Critical Design Features
\thanks{This work was supported by the RESCUE project which has received funding from the European Union's Horizon 2020 research and innovation programme under the Marie Sklodowska-Curie grant agreement No. 722325.}
}

\author{%
\IEEEauthorblockN{%
  Thomas Lange\IEEEauthorrefmark{1}\IEEEauthorrefmark{2},
  Aneesh Balakrishnan\IEEEauthorrefmark{1}\IEEEauthorrefmark{3},
  Maximilien Glorieux\IEEEauthorrefmark{1},
  Dan Alexandrescu\IEEEauthorrefmark{1},
  Luca Sterpone\IEEEauthorrefmark{2}%
}
\IEEEauthorblockA{%
  \IEEEauthorrefmark{1}\textit{iRoC Technologies}, Grenoble, France \\
  \IEEEauthorrefmark{2}\textit{Dipartimento di Informatica e Automatica, Politecnico di Torino}, Torino, Italy \\
  \IEEEauthorrefmark{3}\textit{Department of Computer Systems, Tallinn University of Technology}, Tallinn, Estonia \\
  \{thomas.lange, aneesh.balakrishnan, maximilien.glorieux, dan.alexandrescu\}@iroctech.com \qquad
  luca.sterpone@polito.it}
}

\maketitle

\begin{abstract}

Selective mitigation or selective hardening is an effective technique to obtain a good trade-off between the improvements in the overall reliability of a circuit and the hardware overhead induced by the hardening techniques. Selective mitigation relies on preferentially protecting circuit instances according to their susceptibility and criticality. However, ranking circuit parts in terms of vulnerability usually requires computationally intensive fault-injection simulation campaigns. This paper presents a new methodology which uses machine learning clustering techniques to group flip-flops with similar expected contributions to the overall functional failure rate, based on the analysis of a compact set of features combining attributes from static elements and dynamic elements. Fault simulation campaigns can then be executed on a per-group basis, significantly reducing the time and cost of the evaluation. The effectiveness of grouping similar sensitive flip-flops by machine learning clustering algorithms is evaluated on a practical example.Different clustering algorithms are applied and the results are compared to an ideal selective mitigation obtained by exhaustive fault-injection simulation.

\end{abstract}


\input{01_introduction.tex}

\input{02_background.tex}

\input{03_methodology.tex}

\input{04_results.tex}

\input{05_conclusion.tex}

\bibliographystyle{IEEEtran}
\bibliography{IEEEabrv,bib/VTS_2020.bib}

\end{document}

%% file: 01_introduction.tex
\section{Introduction}

The advancement of the process technologies in the last years made it possible to manufacture chips with tens of millions of flip-flops. At the same time due to the technology scaling, lower supply voltages and higher operating frequencies, circuits became more vulnerable to reliability threats, such as transient faults. Especially, the Soft Errors in flip-flops are a major concern and countermeasures have to be taken into consideration by using hardening techniques, such as Triple Modular Redundancy (TMR). However, a fully protected chip might not meet the system requirements in terms of area, power or target frequency. Since for many applications it is not necessary to decrease the vulnerability to a possible minimum, Selective Mitigation can be used. Thereby, only the most critical elements of the circuit are protected against Soft Errors and thus, the failure rate of the system is decreased to meet all requirements~\cite{polian_scalable_2008, maniatakos_workload-driven_2010, valderas_extensive_2010}.

In order to perform Selective Mitigation an exhaustive failure analysis is required to identify and rank the most vulnerable elements of the circuit. Especially, the failure analysis on a functional level grows with the design size, the number of workloads to analyse and the duration in cycles of each workload. A detailed functional failure analysis requires a significant investment in terms of human efforts, processing resources and tool licenses. Studies have shown that exhaustive fault simulation is not feasible for today's complex circuits~\cite{yu_state_2005}.

\subsection{Objective of This Work}

Identifying and ranking the sequential elements which are most vulnerable to transient faults, usually requires computationally intensive fault-injection simulation campaigns. This procedure can be optimized by grouping flip-flops together which are expected to have a similar sensitivity to faults. Fault injection campaigns can then be performed on a per-group basis and thus, significantly reduce the time and cost of the evaluation~\cite{evans_clustering_2013}. However, this optimization heavily relies on the effectiveness of the grouping methodology. Therefore, we propose a new approach to effectively group flip-flops together which are expected to have a similar sensitivity to functional failures. The approach is based on machine learning clustering techniques which uses a set of features to characterizes each flip-flop in the circuit. The feature set combines attributes from static and dynamic elements. Machine learning clustering algorithms are evaluated and compared to an ideal selective mitigation obtained by exhaustive fault-injection simulation.

\subsection{Organisation of the Paper}

In the following sections, first, the use of clustering techniques for selective mitigation is summarized. Further, the general principle of machine learning clustering techniques are explained. Section~\ref{sec:methodology} presents the proposed methodology to group flip-flops together based on machine learning clustering and the used feature set. The approach is evaluated on a practical example in Section~\ref{sec:results} for different machine learning algorithms. Section~\ref{sec:conclusion} concludes the paper and gives some prospects for future work.

%% file: 02_background.tex
\section{Clustering Techniques for Selective Mitigation}
\label{sec:background}

The approach of protecting only the smallest set of elements in a circuit to meet a specified reliability target is called selective mitigation. Therefore, the individual circuit elements of a circuit need to be ranked from the most to the least sensitive. In the case of transient faults in the sequential logic which lead to a functional failure this usually requires exhaustive fault injection simulation, which might not be feasible for large and complex circuits.

In order to reduce the mentioned fault injection efforts, fault simulation campaigns can be performed on a group basis. Therefore, prior any simulations, flip-flops are grouped together and the statistical fault injection is performed on each of these groups. This can significantly reduce the time and cost of the evaluation. However, the accuracy of this coarse-grained fault injection solely relies on an effective approach to group flip-flops together which have highly similar sensitivity to faults. 

The previous studied techniques for clustering are based on buses, design hierarchy, or a hybrid approach using buses, hierarchy and signal naming information~\cite{evans_clustering_2013}. These approaches have several drawbacks. The bus based and hierarchical based approach are only able to provide a fixed number of clusters. Thereby, the hierarchical based approach often provides a low number of clusters which can be very heterogeneous. The bus based approach often results in a high number of clusters with small number of flip-flops per cluster and one larger cluster containing all the flip-flops which do not belong to any bus. Thus, the reduction of the number of fault injections is limited and the large cluster tends to be heterogeneous which negatively impacts the effectiveness of the clustering. The hybrid approach overcomes the problem of the fixed number of clusters by combining the bus and hierarchy based approaches and also taking the signal names into account. The approach assumes that flip-flops with a similar naming also have a similar function and thus, have a similar sensitivity to faults. However, this relies on a consistent naming convention and strong correlation between the naming and the function within the circuit.

\subsection{Machine Learning Clustering Techniques}

In order to tackle the drawbacks of the previous studied clustering approaches this paper investigates if machine learning clustering techniques are suitable to group the flip-flops. Clustering techniques in the machine learning domain belong to the unsupervised learning category. In general these algorithm try to group similar objects together based on a given set of features. The feature set characterizes the objects and the clustering algorithms group objects together which have similar feature values, while objects from a different group should have highly dissimilar feature values~\cite{alpaydin_introduction_2014}.

\bigskip

In this paper, the flip-flops are characterized by a set of features which will be described in the next section. Machine learning clustering algorithms use this feature set to group flip-flops together. Afterwards, the effectiveness of the grouping will be evaluated on a practical example and it will be determined if flip-flops with similar fault sensitivity are grouped together.

%% file: 03_methodology.tex
\section{Proposed Machine Learning Clustering Approach}
\label{sec:methodology}

The proposed methodology is based on clustering techniques for selective mitigation. In contrary to previous work the clustering approach uses machine learning clustering algorithms and a feature set to characterize each flip-flop in the circuit. In this way no assumptions are made about the design and a general methodology is provided. 

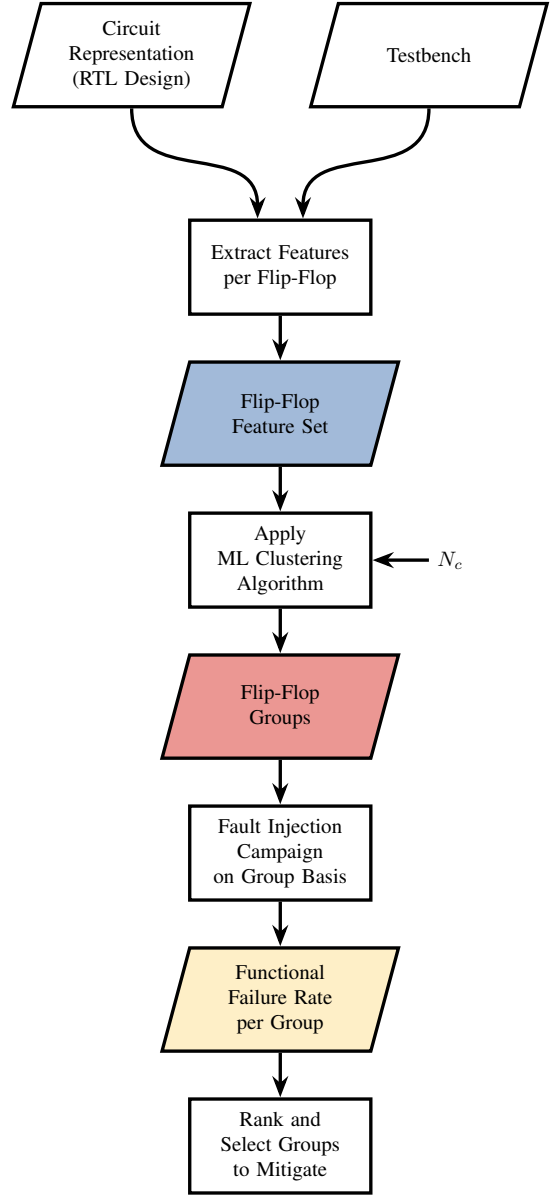
\begin{figure}[htbp!]
    \centering
    \resizebox{0.85\linewidth}{!}{\input{fig/ML-Clustering-Flow.tex}}
    \caption{Selective Mitigation by Using Machine Learning Clustering}
    \label{fig:clustering_flow}
\end{figure}

\input{tab/feature_set.tex}

The steps to perform a selective mitigation are illustrated in Fig.~\ref{fig:clustering_flow}.
First, the features for each flip-flop in the design are extracted by using the RTL description and a corresponding testbench. Second, the machine learning clustering algorithm is applied to the obtained feature set and the flip-flop groups are obtained. The resulting number of groups $N_c$ can be adjusted by the parameters of the machine learning clustering algorithm. The number of groups also dictates the effort needed for the next step: the statistical fault injection. The fault injection campaign is performed on the computed flip-flop cluster and thus, needs more efforts (in terms of computing resources, human efforts and tool licensees) with a higher number of cluster and vice versa. Eventually, the sensitivity to faults for each cluster is obtained and they can be ranked from the most sensitive to the least sensitive. The selective mitigation will be applied starting from the most sensitive cluster until the reliability requirement is met.

\subsection{The Feature Set}

Previous works has shown that the masking effects and the vulnerability of a flip-flop can be related to certain characteristics of the circuit, such as circuit structure and signal probability~\cite{wali_low-cost_2017, ruano_methodology_2009, samudrala_selective_2004}. Motivated by this idea a feature set has been developed in~\cite{lange_machine_2019} which is adapted for the approach presented in this paper. The feature set characterizes each flip-flop instance in the circuit and combines attributes from static elements, such as the circuit structure, as well as dynamic elements, such as the signal activity. 

The original approach was based on the gate-level netlist of a design and contained features which corresponds to synthesis attributes. In order to obtain a more general approach which can already be applied in an early design stage, the methodology presented in this paper uses only features which can be derived from the RTL description of the design (e.g. by performing a fast logic elaboration). Further, two additional features are extracted to reflect the bus and hierarchical based clustering approach described in the previous section. In total the feature set consists of 20 features.

The features extracted for each flip-flop $FF_\text{i}$ are described in detail in Tab.~\ref{tab:feautre_set}. They are divided into two parts, the structural and the signal activity related features. The structural related features describe a flip-flop in relation with other flip-flops in the circuit without taking the (technology dependent) combinatorial logic into account. To consider the workload of the circuit, features are extracted which describe the dynamic behaviour of the flip-flops. Therefore, the information related to the signal activity is considered, such as the state distribution and transitions.

%% file: fig/ML-Clustering-Flow.tex
\newcommand{\boxit}[1]{\parbox[c][9.5ex][c]{7.5em}{\centering #1}}

\pgfdeclarelayer{background}
\pgfsetlayers{background,main}

\begin{tikzpicture}[%
    >={Stealth[scale=1]},              %
    start chain=going below,    %
    node distance=7.5mm and 50mm, %
    every join/.style={norm},   %
    ]
\tikzset{
  base/.style={draw, ultra thick, 
    on chain, on grid, align=center, minimum height=10ex
  },
  proc/.style={base, rectangle, text width=8em},
  term/.style={proc, rounded corners},
  test/.style={base, diamond, aspect=2, text width=5em},
  data/.style={base, trapezium, trapezium stretches=true,
    trapezium left angle=75, trapezium right angle=-75,
    text width=8em,
    node contents={\boxit{#1}}
  },
  coord/.style={coordinate, on chain, on grid, node distance=6mm and 25mm},
  nmark/.style={draw, cyan, circle, font={\sffamily\bfseries}},
  norm/.style={->, ultra thick, draw},
  it/.style={font={\small\itshape}}
}

\node (dsg) [data=Circuit Representation \\ (RTL Design)];
\node (tb) [data=Testbench, right=of dsg];

\node [coord] (flowhmid) at ($(dsg)!0.5!(tb)$) {};
\node [coord] {};

\node [proc, below=20mm of flowhmid] (ext-f) {Extract Features per Flip-Flop};
\node (ff-fset) [data=Flip-Flop \\ Feature Set, text opacity=1.0, 
                fill=set1-darkblue, fill opacity=0.5];
\node [proc] (clustering) {Apply \\ ML Clustering \\ Algorithm};

\node (ff-groups) [data=Flip-Flop \\ Groups, text opacity=1.0, 
                fill=set1-darkred, fill opacity=0.5];
\node [proc] (fi) {Fault Injection Campaign \\ on Group Basis};

\node (fdr) [data=Functional \\ Failure Rate \\ per Group, text opacity=1.0, 
                fill=set1-verylightred, fill opacity=0.5];

\node [proc] (rank) {Rank and \\ Select Groups \\ to Mitigate};

\draw [norm] (dsg) edge[out=-90, in=90, looseness=1.25] (ext-f.115);
\draw [norm] (tb) edge[out=-90, in=90, looseness=1.25] (ext-f.65);

\draw [norm] (ext-f) -- (ff-fset);
\draw [norm] (ff-fset) -- (clustering);
\draw [norm] (clustering) -- (ff-groups);
\draw [norm] (ff-groups) -- (fi);
\draw [norm] (fi) -- (fdr);
\draw [norm] (fdr) -- (rank);

\draw [norm] ($(clustering)+(2.5,0)$) -- (clustering) node[pos=0, right]  {$N_c$};

\end{tikzpicture}

%% file: tab/feature_set.tex
\begin{table*}[htbp!]
    \centering
    \caption{Feature Set to Characterise a Flip-Flop Instance $\text{FF}_i$}
    \label{tab:feautre_set}
    \begin{tabular}{ll%
        p{8.25cm}}
        \toprule
        & \heading{Feature Name} 
            & \heading{Description} \\
        \midrule
        \multicolumn{3}{l}{\textbf{Structural Related Features}} \\
        \cmidrule(lr){1-2}
        & \# FF at Startpoint/Endpoint & %
            The number of flip-flops directly connected to/by $\text{FF}_i$. \\
        & \# Connections from/to FF & %
            The number of flip-flops connected to/by $\text{FF}_i$ within the circuit. \\
        & \# Connections from/to Primary Input/Output & %
            The number of primary inputs/outputs connected to/by $\text{FF}_i$. \\
        & \# FF Stages to/from Primary Input/Output (max/avg/min) & %
            Number of flip-flop stages from/to the primary input/output of the circuit. \\
        & Feedback Depth & %
            The depth (in terms of flip-flops stages) of the shortest feedback loop. \\
        & Bus Position & %
            Describes the position of $\text{FF}_i$ within the bus. \\
        & Bus Length & %
            Describes the total length of the bus signal $\text{FF}_i$ is part of. \\
        & Bus Label & %
            The number/label of the bus signal $\text{FF}_i$ is part of. \\
        & Module Label & %
            The number/label of the hierarchical module $\text{FF}_i$ is part of. \\[5pt]
        \multicolumn{3}{l}{\textbf{Signal Activity Related Features}} \\
        \cmidrule(lr){1-2}
        & @0/@1 & %
            The relative time $\text{FF}_i$ output is at logical~\verb+0+/\verb+1+. \\
        & State Changes & %
            The number of state changes. \\[5pt]
        \bottomrule
    \end{tabular}
\end{table*}

%% file: 04_results.tex
\section{Evaluating Machine Learning Clustering for Selective Mitigation}
\label{sec:results}

In this section the machine learning clustering approach is evaluated on a practical example. Therefore, different clustering algorithms are used to group the flip-flops based on the presented feature set. The effectiveness of the clustering algorithms are measured by evaluating the created flip-flop cluster. The goal is to create flip-flop groups which have a similar vulnerability to critical failures. Therefore, first, an exhaustive full flat statistical fault injection campaign was performed to obtain the sensitivity to critical failures for each flip-flop and to provide an independent measure of the sensitivity. Afterwards, this data is used to evaluate the different clustering algorithms against an ideal and random approach.

\subsection{Circuit Under Test}

For the practical example, the Ethernet 10GE~MAC Core from OpenCores is used. This circuit implements the Media Access Control (MAC) functions as defined in the IEEE~802.3ae standard. The 10GE~MAC core has a 10\,Gbps interface (XGMII TX/RX) to connect it to different types of Ethernet PHYs and one packet interface to transmit and receive packets to/from the user logic~\cite{andre_tanguay_10ge_2013}. The circuit consists of control logic, state machines, FIFOs and memory interfaces. It is implemented at the Register-Transfer Level (RTL) and is publicly available on OpenCores. The RTL description of the design consists of 1054\,flip-flops.

The corresponding testbench writes several packets to the 10GE~MAC transmit packet interface. As packet frames become available in the transmit FIFO, the MAC calculates a CRC and sends them out to the XGMII transmitter. The XGMII~TX interface is looped-back to the XGMII~RX interface in the testbench. The frames are thus processed by the MAC receive engine and stored in the receive FIFO. Eventually, the testbench reads frames from the packet receive interface and prints out the results~\cite{andre_tanguay_10ge_2013}.

\subsection{Full Flat Statistical Fault Injection Results}

In order to evaluate the effectiveness of the clustering algorithms the sensitivity of the considered design was measured by performing a flat statistical fault injection campaign. The simulations are performed at the RT-Level using the corresponding testbench which allows a functional verification. Thus, it is possible to evaluate the system-level impact of errors. The fault injection mechanism consisted of inverting the value stored in a flip-flop by using a simulator functions.

In networking applications, such as the considered design, important data is protected by checksums. This means that a minor payload corruption can be handled by the error correction algorithm. However, in case the fault causes the circuit to stop working and interrupting the flow of sending packages or data is continuously corrupted, then the effect can be considered as critical. Especially, these failures are highly problematic and should be mitigated by selective mitigation.

In each flip-flop 200 faults are injected at a random time during the active phase of the test-case. The functional failure rate of each flip-flop is calculated by dividing the number of simulation runs which lead to a functional failure with the number of total simulation runs. The overall results of the flat statistical fault-injection campaign are presented in Table~\ref{tab:sfi_results}. The average critical failure rate is 5.13\,\% and the most critical flip-flops were identified and ranked.

\begin{table}[htbp!]
  \centering
  \caption{SEU Fault Injection Campaign Results}
  \label{tab:sfi_results}
  \begin{tabular}{lcc}
    \toprule
    & Total & Per Injection \\
    \midrule
    Injection Targets (FFs) & 1054 & - \\
    Injected Faults (SEU) & 210800 & - \\
    Functional Failure & 10814 & 5.13\,\%\\
    \bottomrule
  \end{tabular}
\end{table}

\subsection{Machine Learning Clustering for Selective Mitigation}

The machine learning clustering algorithms are used to group flip-flops based on the feature-set described in section~\ref{sec:methodology}. The features are extracted from the RTL design of the circuit. Therefore, a fast elaboration is performed and the design is converted into a graph representation. The structural features are extracted from the graph by using graph algorithms and the features regarding the signal activity are extracted by tracing the simulation. The feature extraction is automated and overall, takes less than 5 minutes.

The effectiveness of the clustering is evaluated considering they would be used to selectively mitigate against the critical failures. Therefore, the data obtained from the exhaustive statistical fault injection is used to compute the sensitivity to critical faults for each cluster. Then, the reduction of the overall sensitivity of the circuit was calculated by varying the number of groups being protected. For the protection it is assumed that the considered flip-flops within the group are substituted by hardened cells, TMR or other approaches. It is assumed that after mitigation, the sensitivity to Single-Event Upsets is zero\footnote{The authors are aware that this is a simplification and important aspects related to physical design are not considered. However, the here presented approach focuses on the evaluation on the functional level by using the RTL description of the circuit. To obtain a more complete analysis, the presented approach could be combined with e.g. the classical analysis for electrical and temporal masking by using post place and route gate-level netlist.}. The flip-flops to protect were selected starting with the most sensitive clusters first. The results are compared against an ideal and a random approach. In the ideal approach the most sensitive flip-flops are selected based on the exhaustive fault injection campaign. The random approach selects flip-flops to protect randomly (averaged over 100 independent runs).

The clustering algorithms were implemented by using Python's scikit-learn Machine Learning framework~\cite{pedregosa_scikit-learn_2011} and applied to the extracted flip-flop feature set. This process took only several seconds and is negligible. Each considered clustering algorithm has different parameters which can be adjusted. They affect the performance of the clustering and also the resulting number of clusters. For some of the algorithms the number of clusters can be specified directly. Other algorithms try to find the optimal number clusters within the constrained parameters. In the following evaluation the parameters were chosen in a way that the number of resulting clusters are about 20\,\%, 10\,\% and 5\,\% of the number of flip-flops in the circuit. This would correspond to a reduction of the fault injection efforts by $5\times$, $10\times$ and $20\times$ respectively.

\subsubsection{K-Means Clustering}

K-Means clustering aims to partition the given data points into $k$ clusters. The algorithm computes the euclidean distance for each data point in the feature space. The data samples are then separated in such a way the variance within each cluster is equal. The algorithm requires the number of clusters $N_c$ to be specified.

\begin{figure}[htbp!]
    \centering
    \subfloat[From 0\,\% to 100\,\% of mitigated flip-flops]{
        \resizebox{\linewidth}{!}{
            \includegraphics{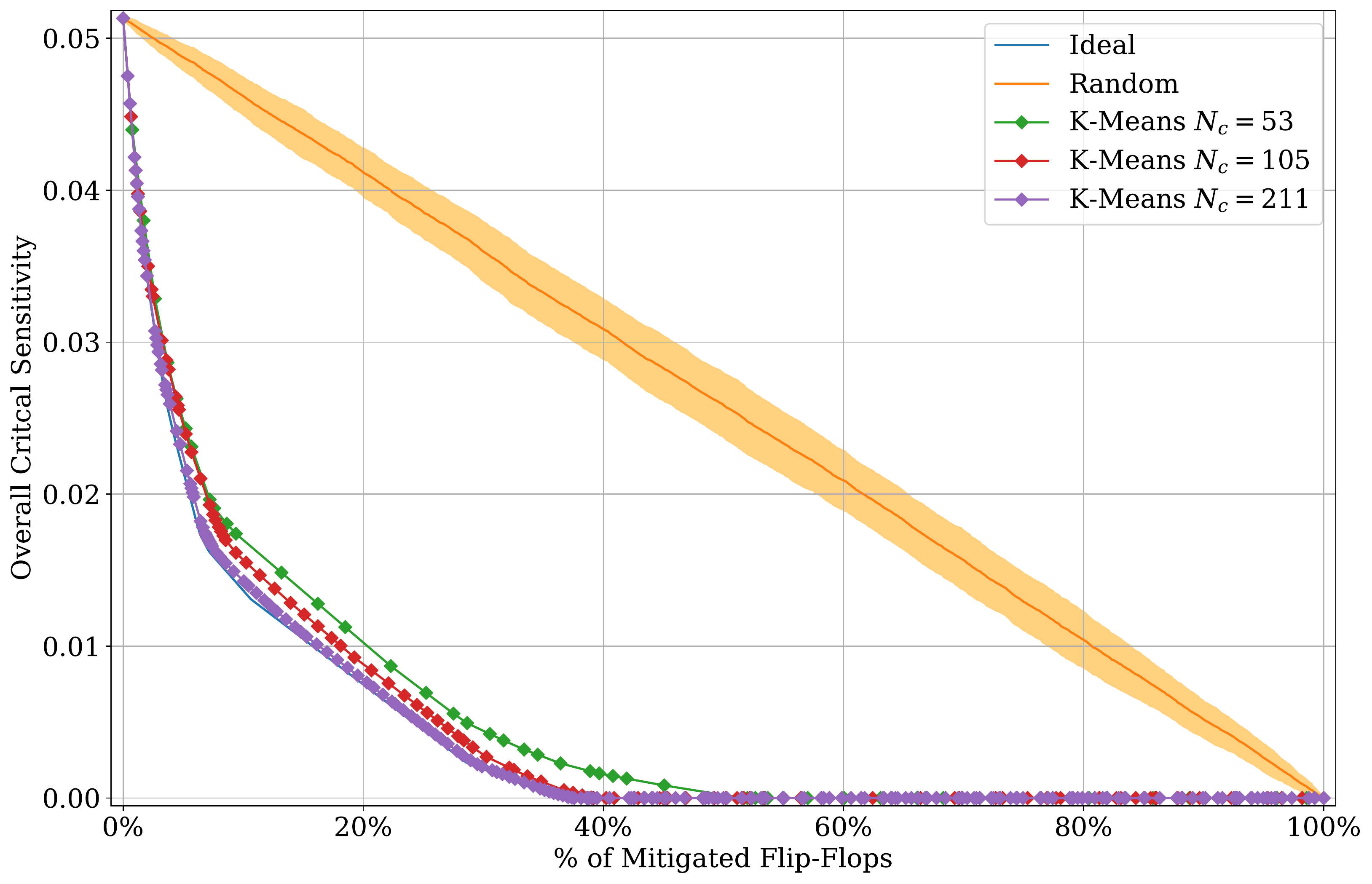}
        }
    }
    
    \subfloat[Section from 0\,\% to 25\,\% of mitigated flip-flops]{
        \resizebox{\linewidth}{!}{
            \includegraphics{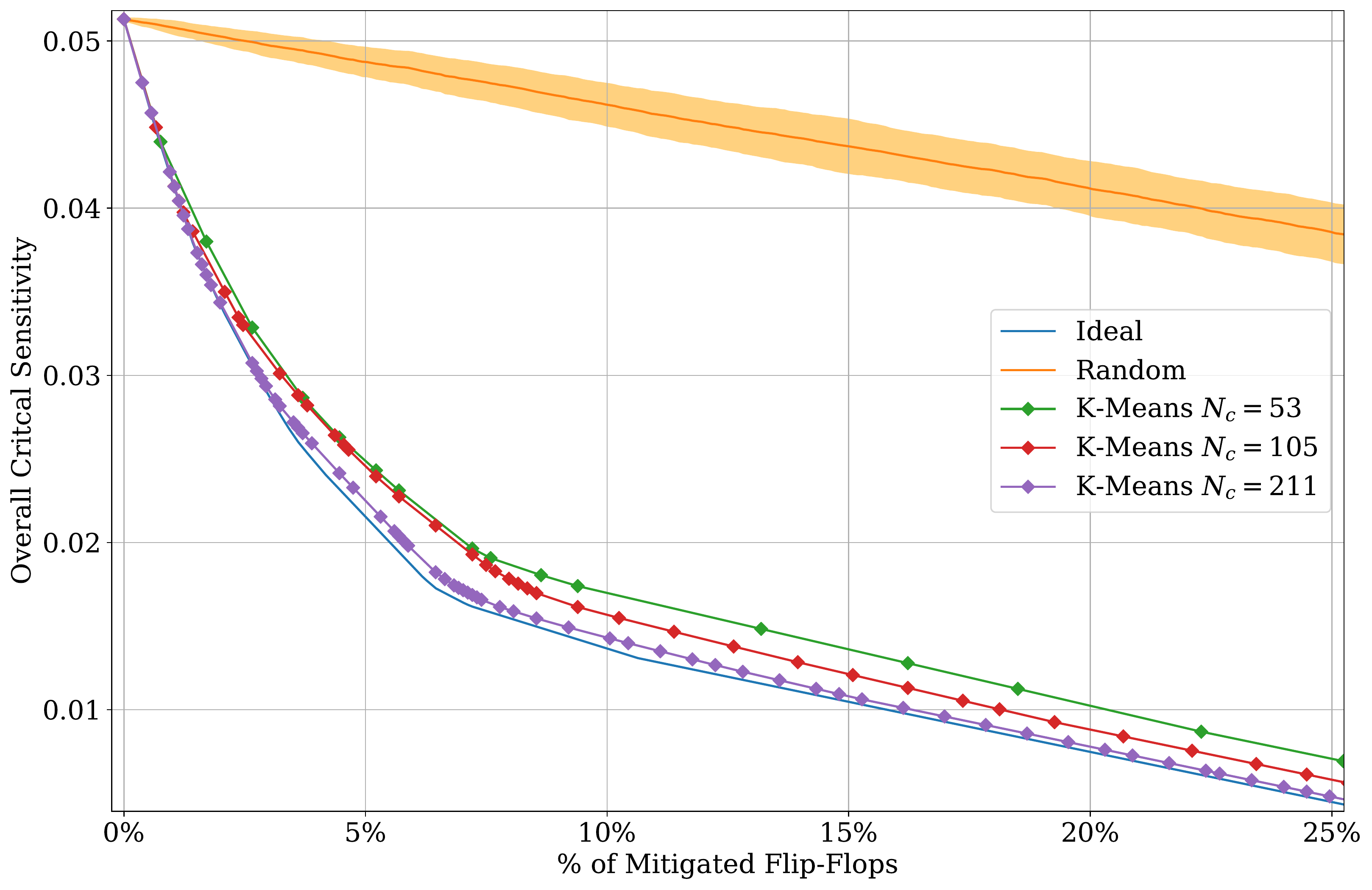}
        }
    }
    \caption{K-Means Clustering}
    \label{fig:kmeans}
\end{figure}

Fig.~\ref{fig:kmeans} shows the overall critical sensitivity when the grouping is performed by the K-Means clustering with different number of clusters $N_c$. The flip-flops to protect were selected starting with the most sensitive clusters first until all flip-flops are mitigated. It can be noted that the grouping performs better when the number of clusters is higher.

\subsubsection{Agglomerative Clustering}

Agglomerative Clustering performs a hierarchical clustering by creating nested clusters, which are represented as a tree. The advantage of hierarchical clustering is that any valid measure of distance can be used, in comparison to K-Means clustering which performs on an euclidean distance metric. Agglomerative Clustering uses a bottom-up approach where each data point starts as its own cluster. Clusters are merged together by following the linkage criteria. This criteria defines the metric used for the merge strategy. It was noted that the best results were obtained by using the maximum or complete linkage, which minimizes the maximum distance between data points of pairs of clusters and a manhatten distance metric (l1 norm).

\begin{figure}[htbp!]
    \centering
    \subfloat[From 0\,\% to 100\,\% of mitigated flip-flops]{
        \resizebox{\linewidth}{!}{
            \includegraphics{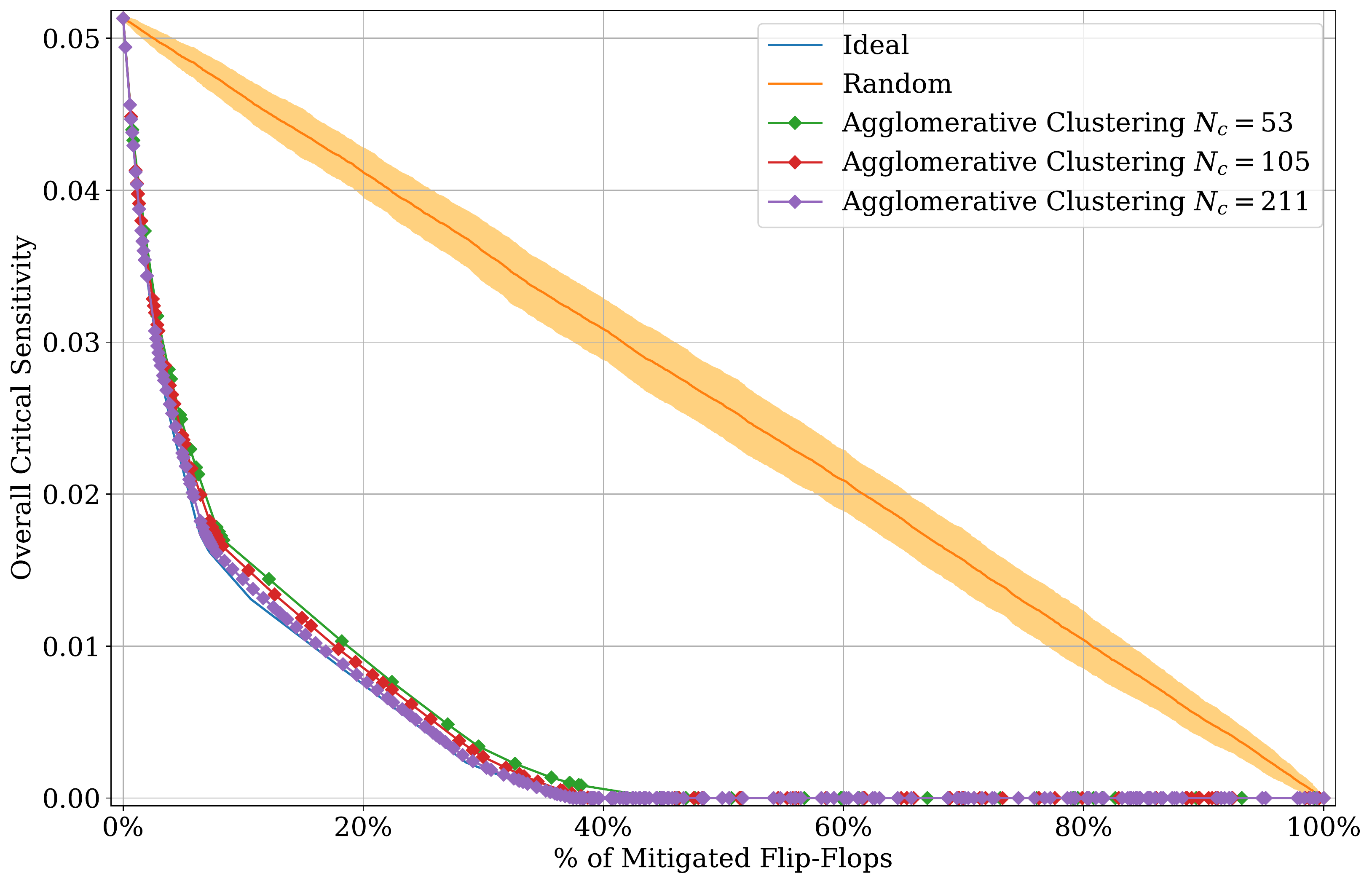}
        }
    }
    
    \subfloat[Section from 0\,\% to 25\,\% of mitigated flip-flops]{
        \resizebox{\linewidth}{!}{
            \includegraphics{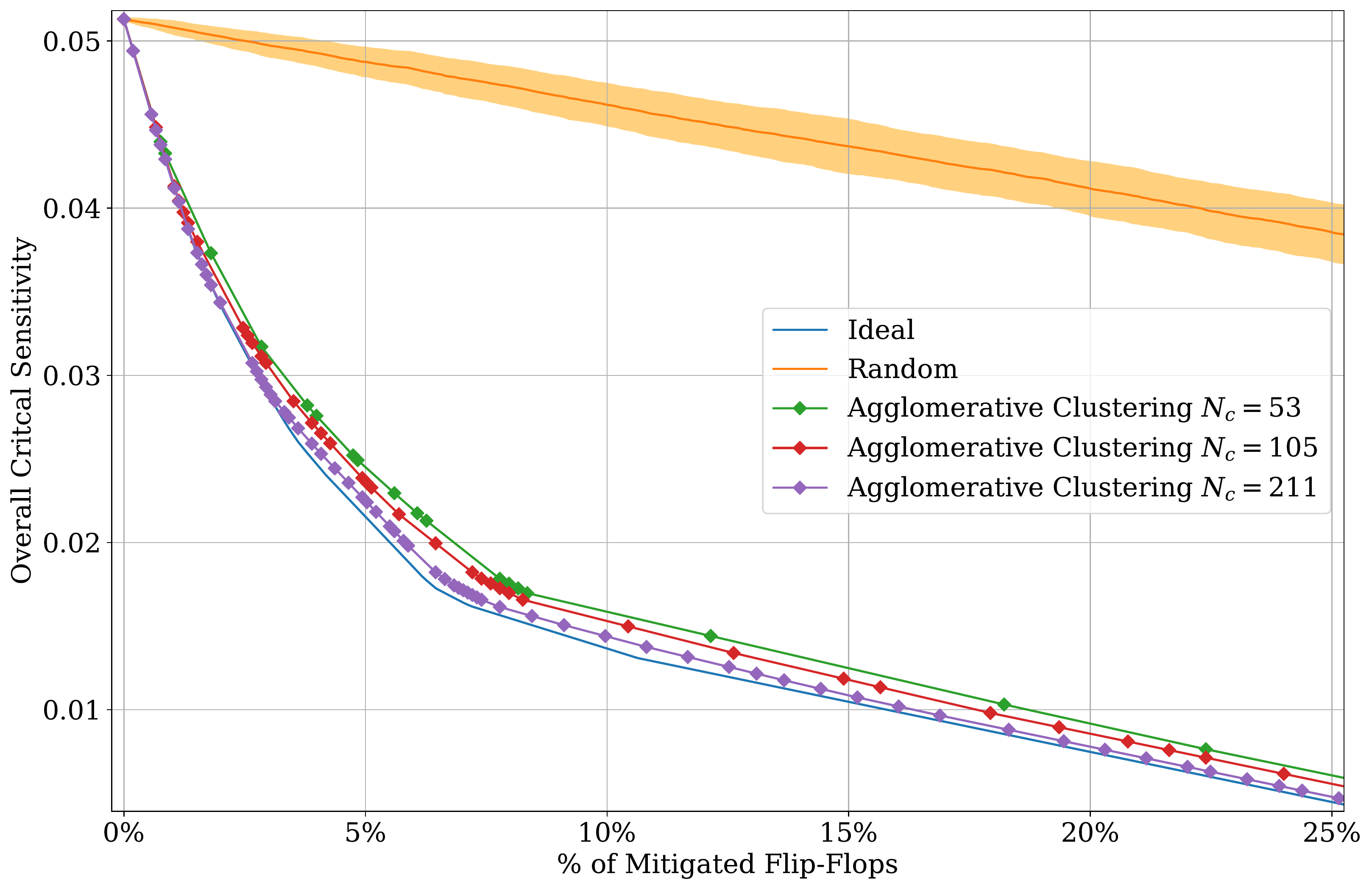}
        }
    }
    \caption{Agglomerative Clustering}
    \label{fig:AgglomerativeClustering}
\end{figure}

In Fig~\ref{fig:AgglomerativeClustering} the results of the selective mitigation are shown when Agglomerative Clustering is used for different number of clusters $N_c$. Similar to the K-Means clustering the effectiveness of the selective mitigation is increasing with a higher number of clusters. However, the improvements from 53 to 105 and 211 clusters are very minor. When comparing the Agglomerative Clustering to K-Means Clustering it can be noted that Agglomerative Clustering performs generally better.

\subsubsection{Mean Shift Clustering}

Mean Shift Clustering aims to find dense areas of data points. The algorithm is based on a sliding-window which is shifted towards regions of higher density. The density of the sliding-window is proportional to the number of data points within the window it. The goal is to locate center points for each group in the dataset. For the previous clustering algorithm the number of clusters had to be specified manually. In Mean Shift clustering the number of clusters is determined by the algorithm. Further, K-Means clustering assumes spherical distribution shape of the clusters in the feature space. For the algorithm the window size $w$ needs to be specified.

\begin{figure}[htbp!]
    \centering
    \subfloat[From 0\,\% to 100\,\% of mitigated flip-flops]{
        \resizebox{\linewidth}{!}{
            \includegraphics{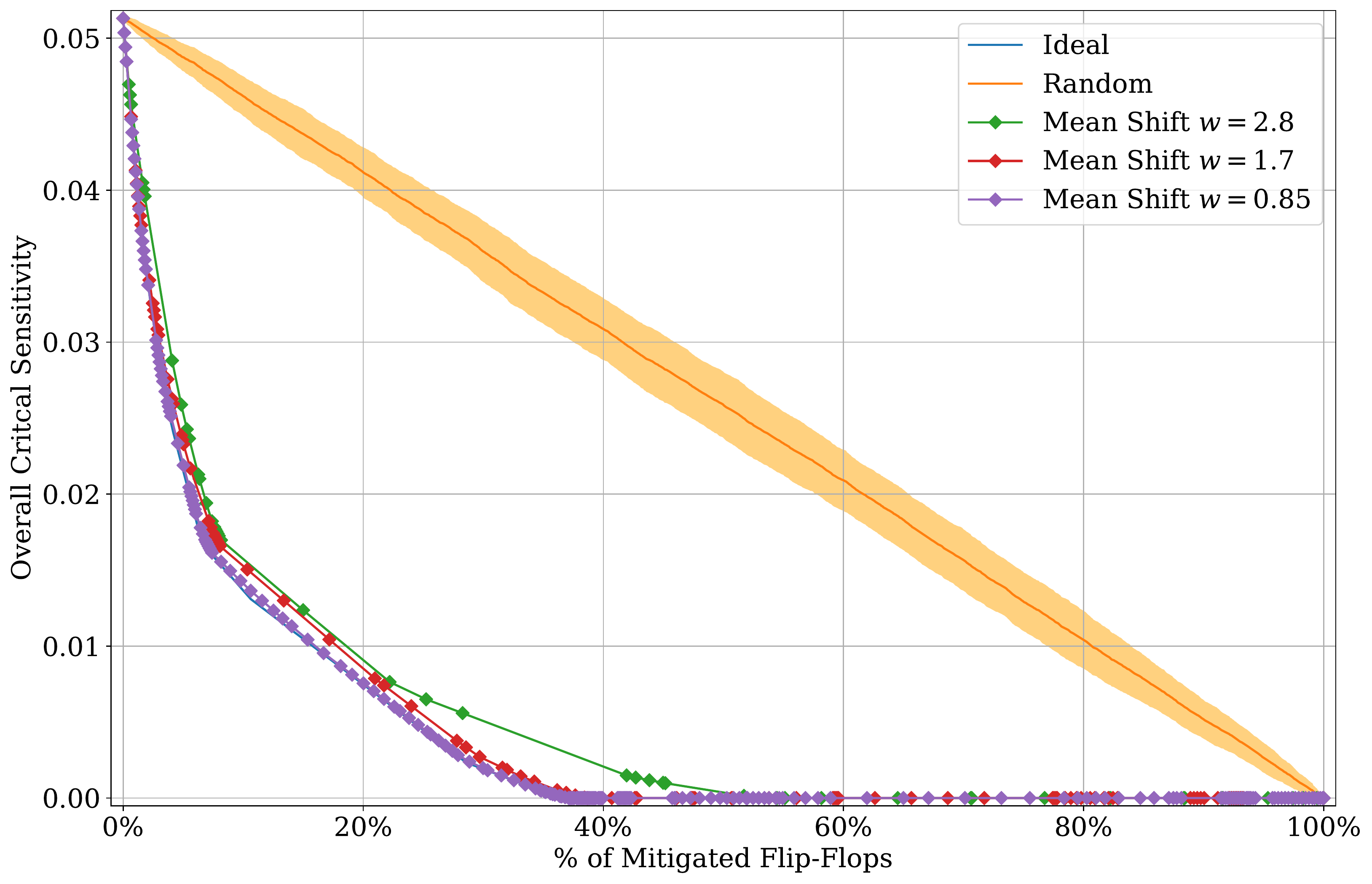}
        }
    }
    
    \subfloat[Section from 0\,\% to 25\,\% of mitigated flip-flops]{
        \resizebox{\linewidth}{!}{
            \includegraphics{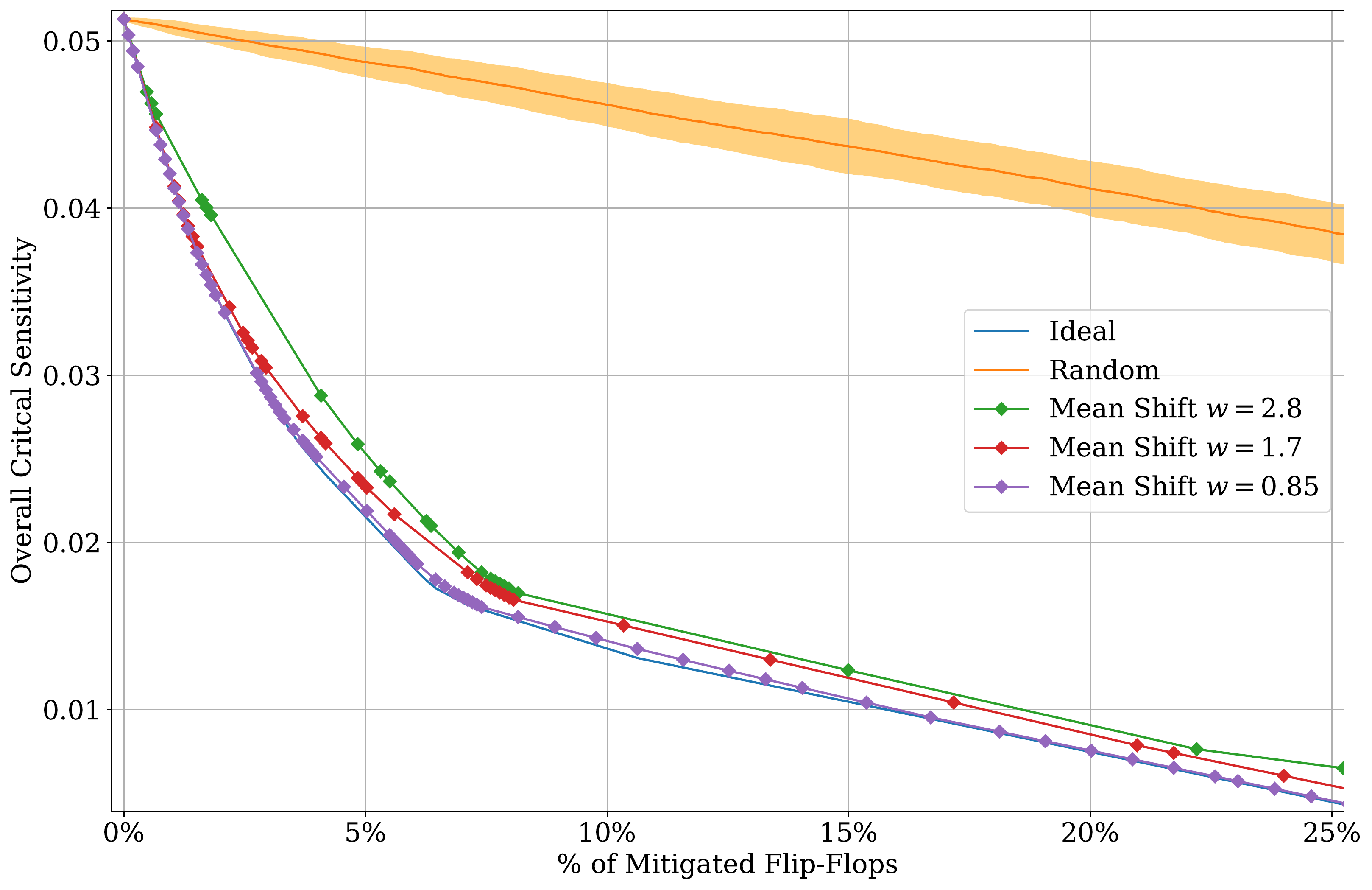}
        }
    }
    \caption{Mean Shift Clustering}
    \label{fig:MeanShift}
\end{figure}

A general problem when using Mean Shift Clustering is to chose the correct windows size. In this analysis the window sizes were chosen in a way the resulting number of clusters are close to the number of clusters used for the previous algorithms. By using window sizes of $w=2.8$, $w=1.7$ and $w=0.85$ the number of clusters were resulting in 52, 105 and 210, respectively. Fig~\ref{fig:MeanShift} shows the effectiveness of the algorithm when using the obtained clusters for selective mitigation. As for the previous algorithms the effectiveness is increasing with a higher number of clusters (smaller window sizes). It can be seen, that the results with a window size of $w=0.85$, which results in 210 groups, is almost as good as the ideal solution.

\subsection{Comparison and Discussion}

The results of the considered clustering algorithms with different resulting number of clusters are summarized in Tab.~\ref{tab:clustering_results}. Since the number of resulting clusters was chosen to be about the same, the average cluster size is identical or very similar. The standard deviation of the cluster sizes however shows that Agglomerative and Mean Shift clustering tend to create cluster with more variety in the cluster size. 

\input{tab/clustering_results}

In order to quantify the effectiveness of the algorithm to create flip-flop groups with similar functional failure rate, two metrics were derived from the results: the average variance of the functional failure rate and the maximum difference of the functional failure rate of flip-flops within the same cluster. An additional metrics was created where the average variance of the functional failure rate within one cluster is weighted with the cluster size. In this way a large cluster, which has the same variance within the cluster as a small cluster, is penalized more.

The results verify what was observed from the Fig.~\ref{fig:kmeans}, \ref{fig:AgglomerativeClustering} and \ref{fig:MeanShift}. In general the algorithms perform better with a higher number of clusters. Agglomerative Clustering performs slightly better than $k$-Means and Mean Shift clustering performs worst with low number of clusters and close to ideal with the highest considered number of clusters.

In order to evaluate the performance of the clustering algorithm without knowledge of the actual/reference values, several metrics exists which quantify certain characteristics of the created clusters (such as the separation of the data). These metrics can be used when the presented approach is applied to a new unknown circuit and different algorithms are evaluated or the algorithm parameters are fine tuned. For this analysis the Davies-Bouldin index was used and results are shown in Tab.~\ref{tab:clustering_results}. This index can be used to evaluate the separation between the clusters. It signifies the average similarity between clusters by comparing the distance between clusters with the size of the clusters themselves. An index of 0 is the lowest possible score and values closer to zero indicate a better partition.

The Davies-Bouldin index does not fully correlate with the functional failure variance or difference metrics. However, it follows the general direction and a lower index can be observed with higher number of clusters. Further, similar scores are obtained for the $k$-Means and Agglomerative Clustering as also seen when comparing the functional failure metrics. The best index, the closest to zero, is also obtained for the best result, Mean Shift clustering with 210 clusters.

%% file: tab/clustering_results.tex
\begin{table*}[htbp!]
    \centering
    \caption{Comparison of the Different Clustering Algorithm}
    \label{tab:clustering_results}
    \begin{tabular}{l ccc ccc c}
        \toprule
        \heading{\shortstack{Clustering \\ Algorithm}}
            & \heading{\# Cluster} 
            & \heading{\shortstack{Mean \\ Cluster Size}}
            & \heading{\shortstack{Standard Deviation \\ Cluster Size}}
            & \heading{\shortstack{Mean \\ Functional Failure \\ Variance}}
            & \heading{\shortstack{Mean Weighted \\ Functional Failure \\ Variance}}
            & \heading{\shortstack{Max  \\ Functional Failure \\ Difference}}
            & \heading{\shortstack{Davies-Bouldin \\ Index}}
            \\
        \midrule
            $k$-Means 
                & 53  & 19.89 & 13.01 & 0.009 & 0.095 & 0.92 & 0.73 \\
                & 105 & 10.04 &  7.45 & 0.010 & 0.041 & 0.92 & 0.68 \\
                & 211 &  4.99 &  3.80 & 0.003 & 0.008 & 0.64 & 0.51 \\[5pt]
            Agglomerative 
                & 53  &	19.89 &	21.39 & 0.015 &  0.094 & 0.92 & 0.78 \\
            Clustering     
                & 105 & 10.04 & 11.76 & 0.011 & 0.040 & 0.92 & 0.57 \\
                & 211 &  4.99 &  5.32 & 0.003 & 0.007 & 0.64 & 0.54 \\[5pt]
            Mean Shift 
                & 52  & 20.27 & 28.71 & 0.017 & 0.167 & 1    & 0.60 \\
                & 105 & 10.04 & 14.12 & 0.006 & 0.038 & 0.92 & 0.42 \\
                & 210 &  5.02 &  6.64 & 0.002 & 0.005 & 0.47 & 0.39 \\
        \bottomrule
    \end{tabular}
\end{table*}

%% file: 05_conclusion.tex
\section{Conclusion and Future Work}
\label{sec:conclusion}

This paper proposes a new methodology to group flip-flops together which are expected to have a similar contributions to the overall functional failure rate. The grouping is based on machine learning clustering and uses a compact set of features combining attributes from static elements and dynamic elements. The advantage in comparison to other existing approaches is that the approach is more flexible and no assumption of the circuit or its representation is made. Further, the number of clusters can be chosen by the user, which determines the needed efforts for the fault injection campaign.

The effectiveness of the grouping by different machine learning clustering algorithms were evaluated on a practical example and compared to an ideal solution. Good results were obtained by choosing the number of clusters with 5\,\% of the total number of flip-flops in the circuit and results close to the ideal solution were obtained with number of clusters corresponding to 10\,\% to 20\,\% of the number of flip-flops. This would mean that the fault injection efforts could be reduced by a factor of $5\times$, $10\times$ or $20\times$ respectively.

Future work will focus on identifying new features to improve the effectiveness of the clustering algorithms as well as applying the techniques to a broader range of circuits. Further, the approach could be extended by using features which take physical design aspects into account and thus, reducing the efforts needed to perform a complete failure analysis on several levels.

%% file: main.bbl
\begin{thebibliography}{10}
\providecommand{\url}[1]{#1}
\csname url@samestyle\endcsname
\providecommand{\newblock}{\relax}
\providecommand{\bibinfo}[2]{#2}
\providecommand{\BIBentrySTDinterwordspacing}{\spaceskip=0pt\relax}
\providecommand{\BIBentryALTinterwordstretchfactor}{4}
\providecommand{\BIBentryALTinterwordspacing}{\spaceskip=\fontdimen2\font plus
\BIBentryALTinterwordstretchfactor\fontdimen3\font minus
  \fontdimen4\font\relax}
\providecommand{\BIBforeignlanguage}[2]{{%
\expandafter\ifx\csname l@#1\endcsname\relax
\typeout{** WARNING: IEEEtran.bst: No hyphenation pattern has been}%
\typeout{** loaded for the language `#1'. Using the pattern for}%
\typeout{** the default language instead.}%
\else
\language=\csname l@#1\endcsname
\fi
#2}}
\providecommand{\BIBdecl}{\relax}
\BIBdecl

\bibitem{polian_scalable_2008}
I.~Polian, S.~M. Reddy, and B.~Becker, ``Scalable {{Calculation}} of {{Logical
  Masking Effects}} for {{Selective Hardening Against Soft Errors}},'' in
  \emph{2008 {{IEEE Computer Society Annual Symposium}} on {{VLSI}}}, Apr.
  2008, pp. 257--262.

\bibitem{maniatakos_workload-driven_2010}
M.~Maniatakos and Y.~Makris, ``Workload-driven selective hardening of control
  state elements in modern microprocessors,'' in \emph{2010 28th {{VLSI Test
  Symposium}} ({{VTS}})}, Apr. 2010, pp. 159--164.

\bibitem{valderas_extensive_2010}
M.~G. Valderas, M.~P. Garcia, C.~Lopez, and L.~Entrena, ``Extensive {{SEU
  Impact Analysis}} of a {{PIC Microprocessor}} for {{Selective Hardening}},''
  \emph{IEEE Transactions on Nuclear Science}, vol.~57, no.~4, pp. 1986--1991,
  Aug. 2010.

\bibitem{yu_state_2005}
Y.~Yu, B.~Bastien, and B.~W. Johnson, ``A state of research review on fault
  injection techniques and a case study,'' in \emph{Annual {{Reliability}} and
  {{Maintainability Symposium}}, 2005. {{Proceedings}}.}, Jan. 2005, pp.
  386--392.

\bibitem{evans_clustering_2013}
A.~Evans, M.~Nicolaidis, S.~Wen, and T.~Asis, ``Clustering techniques and
  statistical fault injection for selective mitigation of {{SEUs}} in
  flip-flops,'' in \emph{International {{Symposium}} on {{Quality Electronic
  Design}} ({{ISQED}})}, Mar. 2013, pp. 727--732.

\bibitem{alpaydin_introduction_2014}
E.~Alpaydin and F.~Bach, \emph{Introduction to {{Machine Learning}}}, ser.
  Adaptive {{Computation}} and {{Machine Learning}} Series.\hskip 1em plus
  0.5em minus 0.4em\relax {MIT Press}, 2014.

\bibitem{wali_low-cost_2017}
I.~Wali, B.~Deveautour, A.~Virazel, A.~Bosio, P.~Girard, and M.~Sonza~Reorda,
  ``\BIBforeignlanguage{en}{A {{Low}}-{{Cost Reliability}} vs. {{Cost
  Trade}}-{{Off Methodology}} to {{Selectively Harden Logic Circuits}}},''
  \emph{\BIBforeignlanguage{en}{Journal of Electronic Testing}}, vol.~33,
  no.~1, pp. 25--36, Feb. 2017.

\bibitem{ruano_methodology_2009}
O.~Ruano, J.~A. Maestro, and P.~Reviriego, ``A {{Methodology}} for {{Automatic
  Insertion}} of {{Selective TMR}} in {{Digital Circuits Affected}} by
  {{SEUs}},'' \emph{IEEE Transactions on Nuclear Science}, vol.~56, no.~4, pp.
  2091--2102, Aug. 2009.

\bibitem{samudrala_selective_2004}
P.~K. Samudrala, J.~Ramos, and S.~Katkoori, ``Selective {{Triple Modular
  Redundancy}} ({{STMR}}) {{Based Single}}-{{Event Upset}} ({{SEU}}) {{Tolerant
  Synthesis}} for {{FPGAs}},'' \emph{IEEE Transactions on Nuclear Science},
  vol.~51, no.~5, pp. 2957--2969, Oct. 2004.

\bibitem{lange_machine_2019}
T.~Lange, A.~Balakrishnan, M.~Glorieux, D.~Alexandrescu, and L.~Sterpone,
  ``Machine {{Learning}} to {{Tackle}} the {{Challenges}} of {{Transient}} and
  {{Soft Errors}} in {{Complex Circuits}},'' in \emph{2019 {{IEEE}} 25th
  {{International Symposium}} on {{On}}-{{Line Testing}} and {{Robust System
  Design}} ({{IOLTS}})}, Jul. 2019, pp. 7--14.

\bibitem{andre_tanguay_10ge_2013}
{Andre Tanguay}, ``{{10GE MAC Core Specification}},'' Jan. 2013.

\bibitem{pedregosa_scikit-learn_2011}
F.~Pedregosa, G.~Varoquaux, A.~Gramfort, V.~Michel, B.~Thirion, O.~Grisel,
  M.~Blondel, P.~Prettenhofer, R.~Weiss, V.~Dubourg, J.~Vanderplas, A.~Passos,
  D.~Cournapeau, M.~Brucher, M.~Perrot, and E.~Duchesnay, ``Scikit-learn:
  {{Machine Learning}} in {{Python}},'' \emph{Journal of Machine Learning
  Research}, vol.~12, pp. 2825--2830, 2011.

\end{thebibliography}
